\documentclass[a4paper,11pt]{article}

\usepackage{jheppub}

\usepackage{graphicx}
\usepackage{multirow}
\usepackage{bm}
\usepackage{braket}
\usepackage{color}
\usepackage{slashed}

\title{ Next-to-next-to-leading-order QCD corrections to double $J/\psi$ production at the $B$ factories }

\author[a,b]{Xu-Dong Huang,}
\emailAdd{huangxd@ihep.ac.cn}

\author[a,b]{Bin Gong,}
\emailAdd{twain@ihep.ac.cn}

\author[a,b]{Rui-Chang Niu,}
\emailAdd{niuruichang@ihep.ac.cn}

\author[c]{Huai-Min Yu,}
\emailAdd{yuhm@stu.pku.edu.cn}

\author[a,b]{Jian-Xiong Wang}
\emailAdd{jxwang@ihep.ac.cn}

\affiliation[a]{Institute of High Energy Physics, Chinese Academy of Sciences, 19B Yuquan Road, Shijingshan District, Beijing, 100049, P.R. China}
\affiliation[b]{University of Chinese Academy of Sciences, Chinese Academy of Sciences, 19A Yuquan Road, Shijingshan District, Beijing, 100049, P.R. China}
\affiliation[c]{School of Physics, Peking University, Beijing 100871, China}

\abstract{In this paper, we study the next-to-next-to-leading-order (NNLO) QCD corrections for the process $e^+e^- \to J/\psi+J/\psi$ at the $B$ factories. By including the NNLO corrections, the cross section turns negative due to the poor convergence of perturbative expansion. Consequently, to obtain a reasonable estimation for the cross section, the square of the amplitude up to NNLO is used. In addition, the contributions from the bottom quark and the light-by-light part, which are usually neglected, are also included. The final cross section is obtained as $1.76^{+2.42}_{-1.66} ~{\rm fb}$ at a center-of-mass energy of $\sqrt{s}=10.58$ GeV. Our result for total cross section and differential cross section could be compared with precise experimental measurement in future at the $B$ factories.
} 

\begin{document}

\maketitle

\flushbottom

\section{Introduction}

In quantum chromodynamics (QCD), the study of heavy quarkonium production assumes an important role in elucidating the interaction between quarks within two-body systems. In processes involving large momentum transfers, perturbative QCD is essential to estimate the theoretical results. In order to apply perturbative QCD to quarkonium production, various models have been introduced, including the color-evaporation model~\cite{Fritzsch:1977ay, Halzen:1977rs}, the color-singlet model~\cite{Chang:1979nn, Berger:1980ni, Matsui:1986dk}, and the nonrelativistic QCD (NRQCD) factorization formalism~\cite{Bodwin:1994jh}. The NRQCD factorization formalism, in particular, allows us to make consistent theoretical predictions and improve them order by order in the QCD coupling constant $\alpha_s$ and the relative velocity of heavy quarks, denoted as $v$.

One of the most intriguing subjects within the domain of heavy quarkonium production and NRQCD is the phenomenon of double charmonium production in $e^+e^-$ annihilation at the $B$ factories. The experimental measurements for the processes $e^+e^- \to J/\psi +\eta_c$ and $e^+e^- \to J/\psi +\chi_{cJ}$ at the $B$ factories have been successfully performed by B{\footnotesize ELLE}~\cite{Belle:2002tfa, Belle:2004abn} and B{\footnotesize A}B{\footnotesize AR}~\cite{BaBar:2005nic}. However, in the case of the process $e^+e^- \to J/\psi +J/\psi$, a clear signal is unable to be detected in B{\footnotesize ELLE}'s measurements, resulting in an upper limit of $\sigma[e^+e^- \to J/\psi +J/\psi] \times \mathcal{B}_{>2} < 9.1~\mathrm{fb}$~\cite{Belle:2004abn}. Here, $\mathcal{B}_{>2}$ signifies the branching ratio of final states involving more than $2$ charged tracks. These experimental findings have sparked considerable theoretical endeavors, with the majority of investigations conducted within the framework of NRQCD factorization. Regarding the processes $e^+e^- \to J/\psi +\eta_c$ and $e^+e^- \to J/\psi +\chi_{cJ}$, their theoretical predictions have been computed up to two-loop level~\cite{Braaten:2002fi, Liu:2002wq, Hagiwara:2003cw, Bodwin:2006ke, He:2007te, Bodwin:2007ga, Zhang:2005cha, Gong:2007db, Dong:2012xx, Li:2013otv, Feng:2019zmt, Huang:2022dfw, Zhang:2008gp, Wang:2011qg, Dong:2011fb, Wang:2013vn, Jiang:2018wmv, Sang:2022kub}. The results exhibit remarkable agreement with the experimental measurements, and further deepen our understanding of these interesting phenomena.

Regarding the process $e^+e^- \to J/\psi +J/\psi$, the calculation at the NRQCD leading order (LO) provides a theoretical prediction for the total cross section about $8.7~\mathrm{fb}$~\cite{Braaten:2002fi}. This value is even greater than the LO NRQCD prediction for the process $e^+e^- \to J/\psi +\eta_c$, and it has been updated to $6.65~\mathrm{fb}$ shortly~\cite{Bodwin:2002kk,Bodwin:2002fk}. With the aid of vector-dominance two-photon exchange model, the authors provide a total cross section about $2.38~\mathrm{fb}$ by exclusively considering the photon fragmentation contribution~\cite{Davier:2006fu}. The non-fragmentation contribution has been investigated in Ref.~\cite{Bodwin:2006yd} within the NRQCD factorization framework, and the authors find a significant destructive interference effect, resulting in a cross section prediction about $1.69\pm0.35~\mathrm{fb}$. The next-to-leading-order (NLO) QCD correction for this process has been studied in Ref.~\cite{Gong:2008ce}, and it is found to be both negative and substantial. Including the NLO correction, the prediction shifts from the range of $7.4\sim9.1~\mathrm{fb}$ to $-3.4\sim2.3~\mathrm{fb}$. The results combining both NLO QCD and relativistic corrections are explored in Ref.~\cite{Fan:2012dy}. The authors find the fixed-order NRQCD prediction for the cross section is in the range of $-12\sim-0.43~\mathrm{fb}$, which is negative and sensitive to the charm quark mass and renormalization scale. They also find that the predicted cross section will shift to the positive range of $1\sim1.5~\mathrm{fb}$ if the approach from Ref.~\cite{Bodwin:2006yd} is used.

In this paper, the next-to-next-to-leading-order (NNLO) QCD corrections to $e^+e^- \to J/\psi+J/\psi$ are studied. It is found that the cross section becomes negative at NNLO, due to the poor convergence of perturbative expansion. In order to handle this, we use the square of the NNLO amplitude to obtain a reasonable result. In addition, the contributions from the bottom quark and light-by-light part, which are usually neglected, are also included. Recently, the NNLO corrections to this process are also studied in Ref.~\cite{Sang:2023liy} where an improved NRQCD approach is proposed. 

The remaining parts of the paper are organized as follows. In Section~\ref{I}, we will provide relevant formulas and offer a concise overview of the calculation. Section~\ref{II} will be dedicated to presenting the numerical results and engaging in discussions. Section~\ref{III} will serve as the summary.

\section{Calculation technology} \label{I}

\subsection{Cross section}

Within the framework of NRQCD factorization, the cross section for $e^+(k_1)e^-(k_2) \to J/\psi(p_1) + J/\psi(p_2)$ can be expressed as follows:
\begin{eqnarray}
d\sigma_{e^+e^- \to J/\psi + J/\psi}=d{\hat \sigma}_{e^+e^- \to (c\bar{c})[n_1]+(c\bar{c})[n_2]}\langle {\cal O}^{J/\psi}(n_1)\rangle\langle {\cal O}^{J/\psi}(n_2)\rangle,
\label{nrqcdfact}
\end{eqnarray}
where $d{\hat \sigma}$ represents the short-distance coefficients (SDCs), $n_{1,2}$ denotes all possible intermediate states, and $\langle {\cal O}^{J/\psi}(n_{1,2})\rangle$ denotes the long-distance matrix elements (LDMEs). In the lowest-order nonrelativistic approximation, only the color-singlet state $^3S_1^{[1]}$ needs to be considered in the summation over $n_1$ and $n_2$, hence we set $n_1=n_2=^3\hspace{-1mm}S_1^{[1]}\equiv n$.

As the LDME $\langle {\cal O}^{J/\psi}(n)\rangle$ incorporates nonperturbative hadronization effects, we initiate our calculation with the cross section of two on-shell $(c\bar{c})$-pairs with the quantum number $^3S_1^{[1]}$. This cross section corresponds to the same SDCs as in $e^+e^-\rightarrow J/\psi +J/\psi$, which can be expressed as:
\begin{eqnarray}
d \sigma_{e^+e^- \to (c\bar{c})[n]+(c\bar{c})[n]}=d{\hat \sigma}_{e^+e^- \to (c\bar{c})[n]+(c\bar{c})[n]}\langle {\cal O}^{(c\bar{c})[n]}(n)\rangle^2. \label{sdcs}
\end{eqnarray}
Here, the symbol $\langle {\cal O}^{(c\bar{c})[n]}(n)\rangle$ is related to NRQCD bilinear operators and can be represented as:
 \begin{eqnarray}
\langle {\cal O}^{(c\bar{c})[n]}(n)\rangle&=&|\langle 0| \chi^{\dagger}{\mathbf \epsilon}\cdot{\mathbf \sigma}\psi|c\bar{c}(n) \rangle|^2,
\end{eqnarray}
where the matrix element $\langle 0| \chi^{\dagger}{\mathbf \epsilon}\cdot{\mathbf \sigma}\psi|c\bar{c}(n) \rangle$ can be computed within the NRQCD framework~\cite{Bodwin:1994jh,Czarnecki:1997vz,Beneke:1997jm,Czarnecki:2001zc,Kniehl:2006qw,Hoang:2006ty,Chung:2020zqc}. On the other hand, the left-hand side of Eq.(\ref{sdcs}) can be directly calculated within perturbative QCD. Consequently, the SDCs $d{\hat \sigma}_{e^+e^- \to (c\bar{c})[n]+(c\bar{c})[n]}$ can be determined from Eq.(\ref{sdcs}). In combination with LDME $\langle {\cal O}^{J/\psi}(n)\rangle$, we can obtain the cross section for $e^+e^- \to J/\psi + J/\psi$ as given in Eq.(\ref{nrqcdfact}).

In the proceeding process, the $e^+ e^-$ pair initially annihilates into two virtual photons, which subsequently turn into two final states. The differential cross section can be conveniently expressed as follows:
\begin{eqnarray}
d \sigma_{e^+e^- \to (c\bar{c})[n]+(c\bar{c})[n]} = \dfrac{1}{4}\dfrac{1}{2s}\sum \vert \mathcal{A} \vert^2 d\Phi_2. \label{dsigma}
\end{eqnarray}
Here, the factor of $1/4$ accounts for the spin average of the initial $e^+e^-$ pair, $1/(2s)$ is the flux factor, $s=(k_1+k_2)^2$ is the squared center-of-mass energy, $\mathcal{A}$ signifies the amplitudes for the process $e^+e^- \to (c\bar{c})[n]+(c\bar{c})[n]$, and $d\Phi_2$ corresponds to the differential phase space for the two-body final state.

\subsection{Calculation of the perturbative SDC}

In this section, we will provide a concise overview of the calculation procedures. Firstly, we utilize the package \texttt{FeynArts}~\cite{Hahn:2000kx} 
to generate corresponding Feynman diagrams and amplitudes for $e^+e^-\to (c\bar{c})[n]+(c\bar{c})[n]$ at NNLO in $\alpha_s$. Secondly, 
we implement the package \texttt{FeynCalc}~\cite{Mertig:1990an,Shtabovenko:2016sxi} to handle Lorentz index contraction and Dirac/$SU(N_c)$ traces. Thirdly, we employ the package \texttt{CalcLoop}\footnote{\texttt{CalcLoop} is a mathematica package developed by Yan-Qing Ma, which is easily accessible at https://gitlab.com/multiloop-pku/calcloop.} to decompose the Feynman amplitudes into 234 Feynman integral families, including the partial fraction decomposition of the linearly-dependent propagators in the integrals. Then the total NNLO amplitudes can be expressed by 87287 Feynman integrals which are further deduced into master integrals by using \texttt{Kira}~\cite{Klappert:2020nbg} (a tool for integration-by-parts reduction). Finally, we utilize the package \texttt{AMFlow}~\cite{Liu:2017jxz, Liu:2020kpc, Liu:2021wks, Liu:2022chg, Liu:2022mfb} to calculate these master integrals.

\begin{figure}[htbp]
\centering
\includegraphics[width=0.8\textwidth]{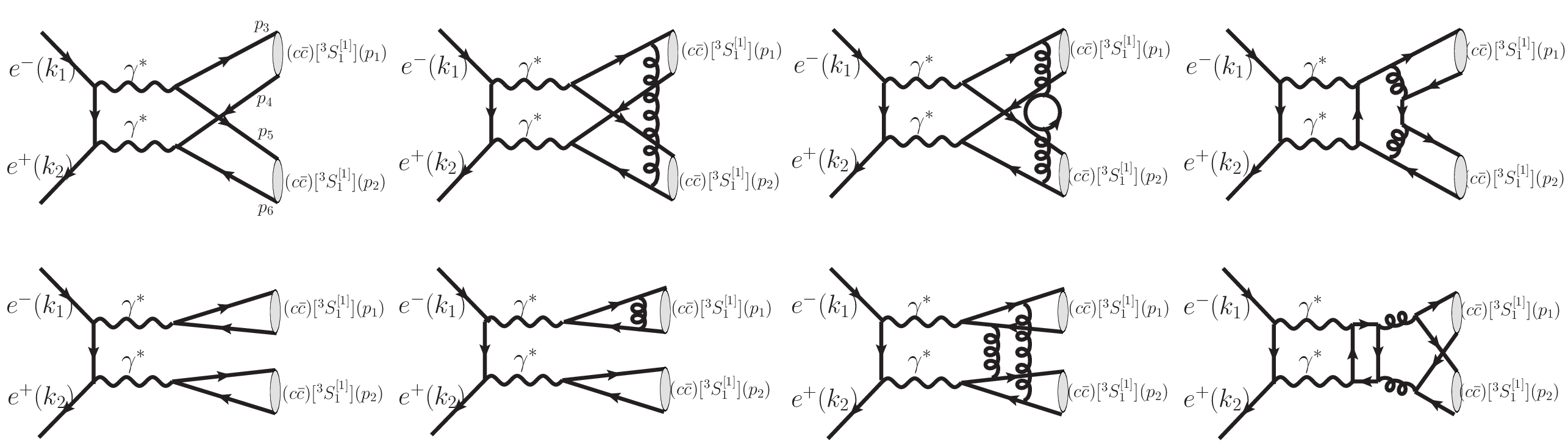}
\caption{Several representative Feynman diagrams for $e^+e^-\to (c\bar{c})[n]+(c\bar{c})[n]$.} \label{feynnlo}
\end{figure}

During the calculation, there are nearly 600 two-loop diagrams for $e^+e^-\to (c\bar{c})[n]+(c\bar{c})[n]$. Several representative Feynman diagrams up to two-loop order are illustrated in Figure~\ref{feynnlo}. Here, $p_3=\frac{p_1+q_1}{2}$, $p_4=\frac{p_1-q_1}{2}$, $p_5=\frac{p_2+q_2}{2}$, and $p_6=\frac{p_2-q_2}{2}$, these momenta $p_{3,4,5,6}$ correspond to the charm and anti-charm quarks and satisfy the on-shell conditions $p_{3,4,5,6}^2=m_c^2$. The momenta $q_1$ and $q_2$ represent the relative momenta between the quark and antiquark within the $c\bar{c}$ pairs. In the lowest-order 
nonrelativistic approximation, the relative momenta $q_1$ and $q_2$ are set to zero, which leads to $p_1^2=p_2^2=(2m_c)^2$. 

We employ the conventional dimensional regularization approach with $d=4-2\epsilon$ to regulate ultraviolet (UV) and infrared (IR) divergences. Feynman diagrams with a virtual gluon line connected to the quark pair in a meson exhibit Coulomb singularity, which manifest as power divergences in the IR limit of relative momentum. Such singularity can be addressed through $c\bar{c}$ wave function renormalization~\cite{Kramer:1995nb,Gong:2007db}. In our calculation, we set the relative momenta $q_1$ and $q_2$ to zero before performing loop integration. The Coulomb divergence vanishes during the calculation with dimensional regularization. The UV divergence is resolved via renormalization. We employ on-shell (OS) renormalization for the heavy quark field and the heavy quark mass. The coupling constant $\alpha_s$ is renormalized in the $\overline{\rm MS}$ scheme. More explicitly, the amplitudes are renormalized according to:
\begin{eqnarray}
\mathcal{A}(\alpha_s,m_Q)=Z_{2,c}^2\bigg[\mathcal{A}_{bare}^{0l}+\mathcal{A}_{bare}^{1l}(\alpha_{s,bare},m_{Q,bare})+\mathcal{A}_{bare}^{2l}(\alpha_{s,bare},m_{Q,bare})\bigg], \label{rn2l1}
\end{eqnarray}
where the $\mathcal{A}^{il}_{bare}|_{i=0,1,2}$ represent the tree, one-loop, and two-loop bare amplitudes, respectively. The $Z_{2,c}$ stands for the on-shell wave-function renormalization constant for the charm quark. The bare mass is renormalized as $m_{Q,bare} = Z_{m,Q} m_Q$, with $Z_{m,Q}$ representing the on-shell mass renormalization constants for heavy quarks. The bare coupling constant is renormalized as:
\begin{eqnarray}
\alpha_{s,bare}=\left(\frac{e^{\gamma_E}}{4\pi}\right)^{\epsilon}\mu_R^{2\epsilon}Z_{\alpha_s}^{\overline{\rm MS}}\alpha_s(\mu_R), \label{alphasbare}
\end{eqnarray}
which corresponds to the $\overline{\rm MS}$ scheme with $n_f$ active flavors. Here, $\mu_R$ represents the renormalization scale, and $Z_{\alpha_s}^{\overline{\rm MS}}$ stands for the renormalization constant of the coupling constant under the $\overline{\rm MS}$ scheme. The renormalization constants $Z_{\alpha_s}^{\overline{\rm MS}}$ up to two-loop level have been presented in Refs.~\cite{Broadhurst:1991fy,Bekavac:2007tk,Czakon:2007ej,Czakon:2007wk,Fael:2020bgs}. Then the renormalized $\mathcal{A}(\alpha_s,m)$ can be obtained by expanding the right-hand side of Eq.~(\ref{rn2l1}) over renormalized quantities up to $\mathcal{O}(\alpha_s^3)$, i.e.,
\begin{eqnarray}
\mathcal{A}(\alpha_s,m_Q)&=&\mathcal{A}^{0l}(m_Q)+\mathcal{A}^{1l}(\alpha_s,m_Q)+\mathcal{A}^{2l}(\alpha_s,m_Q)+\mathcal{O}(\alpha_s^3). \nonumber \\
\end{eqnarray}
Here, the $\mathcal{A}^{il}|_{i=0,1,2}$ represent the tree, one-loop, and two-loop renormalized amplitudes, respectively. The loop integrals are computed with the measure $\mu_R^{2\epsilon}d^dk/(2\pi)^d$, and the corresponding renormalization constants ($Z_{2,c}$, $Z_{m,Q}$, and $Z_{\alpha_s}^{\overline{\rm MS}}$) can be found in Refs.~\cite{Barnreuther:2013qvf,Tao:2022qxa}. Thus, the differential cross section can be obtained as:
\begin{eqnarray}
\frac{d\sigma_{e^+e^- \to (c\bar{c})[n]+(c\bar{c})[n]}}{d\vert\cos\theta\vert}&=&\frac{1}{8 s}\frac{\kappa}{16\pi}\bigg|\mathcal{A}^{0l}+\mathcal{A}^{1l}+\mathcal{A}^{2l}+\mathcal{O}(\alpha_s^3)\bigg|^2 \nonumber \\
&=&\frac{1}{8 s}\frac{\kappa}{16\pi} \Big[\vert \mathcal{A}^{0l}\vert ^2+2{\rm Re}(\mathcal{A}^{1l}\mathcal{A}^{0l,*})+2{\rm Re}(\mathcal{A}^{2l}\mathcal{A}^{0l,*})+\vert \mathcal{A}^{1l}\vert ^2 \nonumber \\[1mm]
&&+2{\rm Re}(\mathcal{A}^{2l}\mathcal{A}^{1l,*})+\vert \mathcal{A}^{2l}\vert ^2+\cdots\Big], \label{ampsqrt}
\end{eqnarray}
where $\kappa=\sqrt{1-(16m_c^2)/s}$ and $\theta$ is the angle between the $J/\psi$ and the beam.

To calculate the square of NNLO amplitude as shown in Eq.~(\ref{ampsqrt}), we try to obtain the amplitude $\mathcal{A}^{nl}|_{n=0,1,2}$ with the help of a complete basis space. These bases are constructed based on the Lorentz structures governing the process $e^+e^-\to (c\bar{c})[n]+(c\bar{c})[n]$. They are obtained as
\begin{eqnarray}
\left(
\begin{array}{ccc}
  \vert e_1 \rangle \\
  \vert e_2 \rangle \\
  \vert e_3 \rangle \\
  \vert e_4 \rangle \\
  \vert e_5 \rangle \\
  \vert e_6 \rangle \\
  \vert e_7 \rangle \\
  \vert e_8 \rangle \\
  \vert e_9 \rangle \\
  \vert e_{10} \rangle 
\end{array}
\right)=
\left(
\begin{array}{ccc}
  g^{\rho_1 \rho_2}\bar{v}(k_2)\slashed{p}_2u(k_1) \\
  k_1^{\rho_1}k_1^{\rho_2}\bar{v}(k_2)\slashed{p}_2u(k_1)+k_2^{\rho_1}k_2^{\rho_2}\bar{v}(k_2)\slashed{p}_2u(k_1) \\
  k_1^{\rho_1}k_1^{\rho_2}\bar{v}(k_2)\slashed{p}_2u(k_1)-k_2^{\rho_1}k_2^{\rho_2}\bar{v}(k_2)\slashed{p}_2u(k_1) \\
  k_1^{\rho_1}k_2^{\rho_2}\bar{v}(k_2)\slashed{p}_2u(k_1) \\
  k_2^{\rho_1}k_1^{\rho_2}\bar{v}(k_2)\slashed{p}_2u(k_1) \\
  k_1^{\rho_2}\bar{v}(k_2)\gamma^{\rho_1}u(k_1)+k_2^{\rho_1}\bar{v}(k_2)\gamma^{\rho_2}u(k_1) \\
  k_1^{\rho_2}\bar{v}(k_2)\gamma^{\rho_1}u(k_1)-k_2^{\rho_1}\bar{v}(k_2)\gamma^{\rho_2}u(k_1) \\
  k_1^{\rho_1}\bar{v}(k_2)\gamma^{\rho_2}u(k_1)+k_2^{\rho_2}\bar{v}(k_2)\gamma^{\rho_1}u(k_1) \\
  k_1^{\rho_1}\bar{v}(k_2)\gamma^{\rho_2}u(k_1)-k_2^{\rho_2}\bar{v}(k_2)\gamma^{\rho_1}u(k_1) \\
\bar{v}(k_2)\slashed{p}_2\gamma^{\rho_1}\gamma^{\rho_2}u(k_1) 
\end{array}
\right).
\end{eqnarray}
Here, $\rho_1$ and $\rho_2$ represent the Lorentz indices of the two final states, $u$ and $\bar{v}$ are the Dirac spinors of initial lepton pair in which spin notation is omitted, and the relationship $p_1=k_1+k_2-p_2$ has been applied. It is worth noting that all the external particles in the process are colorless. Therefore, no color projection operator appears in the amplitude basis. This allows us to express the amplitude as follows:
\begin{eqnarray}
\mathcal{A}^{nl}|_{n=0,1,2}=\sum_{i=1}^{10} c_i^{nl} \vert e_{i} \rangle.
\end{eqnarray}
The coefficients $c_i$ are determined by
\begin{eqnarray}
c_i^{nl}|_{n=0,1,2}=\sum_{j=1}^{10} G_{i,j}^{-1} d_j^{nl},
\end{eqnarray}
where $d_i^{nl}|_{n=0,1,2}=\langle e_{i} \vert \mathcal{A}^{nl} $ is the inner product of the amplitude $\mathcal{A}^{nl}$ and the basis $\vert e_{i} \rangle$, and $G_{i,j}= \langle e_{j} \vert e_{i} \rangle$ signifies the inner product of the basis $\vert e_{i} \rangle$ and the basis $\vert e_{j} \rangle$. It has been observed that the coefficients $d_3$, $d_6$, and $d_8$ are zero at the tree, one-loop, and two-loop levels, respectively. Then the products $\mathcal{A}^{ml}\mathcal{A}^{nl,*}$ can be obtained using the following expression:
\begin{eqnarray}
\mathcal{A}^{ml}\mathcal{A}^{nl,*}=\sum_{i=1}^{10}\sum_{j=1}^{10} c_i^{ml} G_{i,j} c_j^{nl,*},
\end{eqnarray}
where $m$ and $n$ take on values of 0, 1, and 2, respectively.

However, there are remaining IR divergences in $\mathcal{A}^{2l}$. This further makes $\mathcal{A}^{2l}\mathcal{A}^{0l,*}$, $\mathcal{A}^{2l}\mathcal{A}^{1l,*}$, and $\vert \mathcal{A}^{2l}\vert ^2$ divergent. On the other hand, as already known, $\langle {\cal O}^{(c\bar{c})[n}(n)\rangle$ also becomes IR divergent at two-loop level. In the leading order of $v^2$ and within the $\overline{\rm MS}$ scheme, it can be expressed as 
\begin{eqnarray}
\langle {\cal O}^{(c\bar{c})[n]}(n)\rangle|_{\overline{\rm MS}}&=&2N_c\bigg[1-\alpha_s^2(\mu_R)\left(\frac{\mu_{\Lambda}^2e^{\gamma_E}}{\mu_R^2 4\pi}\right)^{-2\epsilon}\left(\frac{C_F^2}{3}+\frac{C_F C_A}{2}\right)\frac{1}{2\epsilon}\bigg]. \label{ldscc}
\end{eqnarray}
which is derived from Refs.~\cite{Bodwin:1994jh,Czarnecki:1997vz,Beneke:1997jm,Czarnecki:2001zc,Kniehl:2006qw,Hoang:2006ty,Chung:2020zqc}.
Here the term $\left(\mu_{\Lambda}^2/\mu_R^2\right)^{-2\epsilon}$ arises from the evolution of the $\alpha_s$, from the factorization scale $\mu_{\Lambda}$ to the renormalization scale $\mu_R$, since the correction is initially obtained at the scale $\mu_{\Lambda}$. The factor $[e^{\gamma_E}/(4\pi)]^{-2\epsilon}$ is a consequence of the $\alpha_s$ definition within the $\overline{\rm MS}$ scheme, as given in Eq.~(\ref{alphasbare}). 
The divergence found in $\mathcal{A}^{2l}\mathcal{A}^{nl,*}$ is exactly same as this one, which renders the SDCs obtained from Eq.(\ref{sdcs}) without any divergences.
Meanwhile this introduces an explicit logarithmic dependence on the NRQCD factorization scale $\mu_{\Lambda}$, which is on the order of $\ln({\mu_\Lambda^2}/{m^2})$ in the SDCs. On the other hand, this $\mu_{\Lambda}$ dependence can be completely cancelled by considering the $\mu_{\Lambda}$ dependence of the LDMEs at fixed order.

The LDME $\langle {\cal O}^{J/\psi}(n)\rangle$ is often approximated as follows:
\begin{eqnarray}
\langle {\cal O}^{J/\psi}(n)\rangle\approx N_c \vert R_s^{J/\psi}(0)\vert^2/(2\pi),
\end{eqnarray}
where $R_s^{J/\psi}(0)$ represents the wave function of $J/\psi$ at the origin. Combining this approximation with Eqs.~(\ref{nrqcdfact}, \ref{sdcs}, \ref{ampsqrt}, \ref{ldscc}), the differential cross section for $e^+e^- \to J/\psi + J/\psi$ can be expressed in the following form:
\begin{eqnarray}
\frac{d\sigma_{e^+e^- \to J/\psi + J/\psi}}{d\vert\cos\theta\vert}&=&\frac{d\sigma_{e^+e^- \to (c\bar{c})[n]+(c\bar{c})[n]}}{d\vert\cos\theta\vert}\frac{\langle {\cal O}^{J/\psi}(n)\rangle^2}{\langle {\cal O}^{(c\bar{c})[n]}(n)\rangle^2|_{\overline{\rm MS}}} \nonumber \\
&=&(f_{0}+f_{1}\alpha_s+f_{2}\alpha_s^2+f_{3}\alpha_s^3+f_{4}\alpha_s^4+\cdots)\vert R_s^{J/\psi}(0)\vert^4. \label{dsigmadcos}
\end{eqnarray}
Here, the terms $f_{i}|_{i=0,1,2,3,4}$ represent the SDCs at corresponding perturbative orders. It should be noted that the results for $f_{3}$ and $f_{4}$ are divergence free and gauge invariant, but incomplete, since only the contributions from $\mathcal{A}^{2l}\mathcal{A}^{1l,*}$ and $\vert \mathcal{A}^{2l}\vert ^2$ are considered here.

\section{Phenomenological results} \label{II}

For the numerical calculations, we use the following input parameters:
\begin{eqnarray}
m_b=4.8 {\rm GeV}, \alpha_s(m_{_Z})=0.1179, \sqrt{s}=10.58 {\rm GeV}, \alpha(2m_c)=1/132.6. 
\end{eqnarray}
Here, the bottom quark pole mass and running QCD coupling constant at the scale $m_{_Z}$ are taken from Particle Data Group~\cite{ParticleDataGroup:2022pth}, and the QED coupling constant at the scale $2m_c$ is taken from Refs.~\cite{Bodwin:2007ga,Sang:2023liy}. $m_{_Z}=91.1876$ GeV is the $Z$ boson mass and $m_c$ is the charm quark pole mass. We use the package \texttt{RunDec3}~\cite{Herren:2017osy} to evaluate the running QCD coupling constant $\alpha_s(\mu_R)$ at three-loop accuracy.

The value of $R_s^{J/\psi}(0)$ can be extracted from the leptonic decay width at the two-loop level using the following formula~\cite{Kallen:1955fb,Beneke:1997jm,Czarnecki:1997vz,Kniehl:2006qw,Egner:2021lxd}:
\begin{eqnarray}
\Gamma_{J/\psi \to e^+e^-}&=&\frac{4\alpha^2e_c^2}{m^2_{J/\psi}}\vert R_s^{J/\psi}(0)\vert^2\bigg\{1-2C_F\frac{\alpha_s}{\pi}+\left(\frac{\alpha_s}{\pi}\right)^2\bigg[-2C_F\tilde{\beta}_0\ln\frac{\mu_R^2}{m_c^2}-3\pi^2C_F\bigg(\frac{1}{18}C_F \nonumber \\
&&+\frac{1}{12}C_A\bigg)\ln\frac{\mu_\Lambda^2}{m_c^2}+C_AC_F\bigg(\frac{89\pi^2}{144}-\frac{151}{72}-\frac{5\pi^2}{6}\ln2
-\frac{13}{4}\zeta_3\bigg)\nonumber \\
&&+C_F^2\bigg(\frac{23}{8}-\frac{79\pi^2}{36}+\pi^2\ln2-\frac{1}{2}\zeta_3\bigg)+C_FT_Fn_H\left(\frac{22}{9}-\frac{2\pi^2}{9}\right)\nonumber \\
&&+\frac{11}{18}C_FT_Fn_L+n_Mc_b\bigg]\bigg\}^2, \label{gammajpsi}
\end{eqnarray}
where $e_c$, $C_F$, $C_A$, $T_F$, $\zeta_3$, and $\tilde{\beta}_0=\left(11-2n_f/3\right)/4$ represent various constants and parameters. For the fermion-loop contributions to $\Gamma_{J/\psi \to e^+e^-}$, it is convenient to introduce $n_f=n_L+n_H+n_M$. Here $n_L=3$ counts the massless quarks, $n_H=1$ and $n_M=1$ label the contributions with closed massive quark loops with charm quark and bottom quark, respectively. The full expression of $c_b$ is shown in Ref.~\cite{Egner:2021lxd}, whose numerical value is $-0.394$ in present case. By choosing $\Gamma_{J/\psi \to e^+e^-}=5.53 {\rm KeV}$~\cite{ParticleDataGroup:2022pth}, $m_{J/\psi}=2m_c$, $\mu_R=2m_c$, and $m_c=1.5 {\rm GeV}$, we obtain the values of $\vert R_s^{J/\psi}(0)\vert^2$, which are 0.492 GeV$^3$ at the tree level, 0.796 GeV$^3$ at the one-loop level, and $\vert R_s^{J/\psi}(0)\vert^2_{\mu_{\Lambda}=1 {\rm GeV}}$ as 1.810 GeV$^3$ at the two-loop level\footnote{The leptonic decay width, denoted as $\Gamma_{J/\psi \to e^+e^-}$, is related to the decay constant $f_{J/\psi}$ through the relationship: $\Gamma_{J/\psi \to e^+e^-}=(4\pi\alpha^2)/(3m^2_{J/\psi})|f_{J/\psi}|^2$. As the same treatment as taken in Refs.~\cite{Beneke:1997jm,Bodwin:2007fz,Feng:2022vvk}, the leptonic decay widths at the $n$-loop level correspond to the calculation of the decay constant $f_{J/\psi}$ up to the $n$-loop level. Consequently, the predictions for the leptonic decay widths are essentially the square of the decay constant $f_{J/\psi}$. If we derive the value of $R_s^{J/\psi}(0)$ using the expression of the leptonic decay widths up to $\alpha_s^2$-order, the numerical result for $\vert R_s^{J/\psi}(0)\vert^2=5.528$ GeV$^3$ is much bigger than the values estimated in various theoretical models~\cite{Eichten:1995ch,Choe:2003wx,Gray:2005ur,Rai:2008sc,Azhothkaran:2020ipl,Akbar:2015evy,Akbar:2011jd,Radford:2007vd,Bodwin:2007fz,Egner:2021lxd,Beneke:2014qea}. A collection of the values of $\vert R_s^{J/\psi}(0)\vert^2$ are given in Ref.~\cite{Feng:2022vvk}, which shows the $\vert R_s^{J/\psi}(0)\vert^2=0.810$ GeV$^3$ in the Buchm\"{u}ller-Tye potential model, $\vert R_s^{J/\psi}(0)\vert^2=1.1184$ GeV$^3$ in the Lattice NRQCD, and $\vert R_s^{J/\psi}(0)\vert^2=1.454$ GeV$^3$ in the Cornell potential model. This discrepancy is due to the poor convergence of perturbative expansion in the leptonic decay width $\Gamma_{J/\psi \to e^+e^-}$. However, this discrepancy can be mitigated by employing Eq.~(\ref{gammajpsi}), which includes certain higher-order corrections in $\alpha_s$. In a similar vein, we study the square of NNLO amplitude for the process $e^+e^- \to J/\psi + J/\psi$ to obtain a reasonable theoretical prediction.}. In the following, we consider the differential cross section with eleven sample points of $\vert\cos\theta\vert$ ranging from 0 to 1.

\begin{table}\tiny
\centering
\begin{tabular}{c c c c}
\hline
$\vert\cos\theta\vert$ & $f_{0}$ & $f_{1}$ & $f_{2}$ \\
\hline
$0.193$ & $3.0687$ & $-~11.1472$ & $0.5647n_L+0.2162n_H-1.3447n_M-0.0320lbl-41.1089-11.1472\beta_0L_{\mu}-15.9116L_{\mu_{\Lambda}}$ \\
\hline
$0.402$ & $3.8973$ & $-~14.2469$ & $0.7247n_L+0.2698n_H-1.7184n_M-0.0344lbl-51.9534-14.2469\beta_0L_{\mu}-20.2080L_{\mu_{\Lambda}}$ \\
\hline
$0.601$ & $5.9069$ & $-~21.6244$ & $1.1036n_L+0.4028n_H-2.6079n_M-0.0386lbl-78.6394-21.6244\beta_0L_{\mu}-30.6282L_{\mu_{\Lambda}}$ \\
\hline
$0.698$ & $7.9392$ & $-~28.9429$ & $1.4775n_L+0.5403n_H-3.4904n_M-0.0416lbl-105.9859-28.9429\beta_0L_{\mu}-41.1664L_{\mu_{\Lambda}}$ \\
\hline
$0.800$ & $12.0746$ & $-~43.5649$ & $2.2221n_L+0.8261n_H-5.2535n_M-0.0459lbl-162.2353-43.5649\beta_0L_{\mu}-62.6088L_{\mu_{\Lambda}}$ \\
\hline
$0.849$ & $15.7238$ & $-~56.2870$ & $2.8694n_L+1.0826n_H-6.7876n_M-0.0487lbl-212.2457-56.2870\beta_0L_{\mu}-81.5310L_{\mu_{\Lambda}}$ \\
\hline
$0.902$ & $22.8893$ & $-~80.9980$ & $4.1287n_L+1.5932n_H-9.7673n_M-0.0529lbl-310.9459-80.9980\beta_0L_{\mu}-118.6851L_{\mu_{\Lambda}}$ \\
\hline
$0.922$ & $27.1569$ & $-~95.6123$ & $4.8758n_L+1.9006n_H-11.5295n_M-0.0551lbl-369.9169-95.6123\beta_0L_{\mu}-140.8136L_{\mu_{\Lambda}}$ \\
\hline
$0.951$ & $37.0190$ & $-129.2083$ & $6.6029n_L+2.6190n_H-15.5804n_M-0.0600lbl-506.5262-129.2083\beta_0L_{\mu}-191.9502L_{\mu_{\Lambda}}$ \\
\hline
$0.975$ & $50.9428$ & $-176.3416$ & $9.0683n_L+3.6601n_H-21.2633n_M-0.0678lbl-700.1246-176.3416\beta_0L_{\mu}-264.1479L_{\mu_{\Lambda}}$ \\
\hline
$0.999$ & $54.7376$ & $-187.8744$ & $10.2369n_L+4.2634n_H-22.6496n_M-0.0923lbl-758.7978-187.8744\beta_0L_{\mu}-283.8247L_{\mu_{\Lambda}}$ \\
\hline
\end{tabular}
\caption{The SDCs $f_{i}|_{i=0,1,2}$ for eleven different values of $\vert\cos\theta\vert$. Here, $\beta_0=\frac{1}{4\pi}\left(11-\frac{2}{3}n_f\right)$, $L_{\mu}=\ln\frac{\mu_R^2}{m_c^2}$, and $L_{\mu_{\Lambda}}=\ln\frac{\mu_{\Lambda}^2}{m_c^2}$. In our results, we set $n_f=5$, $n_L=3$, $n_H=1$, $n_M=1$, and $lbl=1$. }  \label{SDCsfi1}
\end{table}

In Tables~\ref{SDCsfi1},~\ref{SDCsfi2}, and~\ref{SDCsfi3}, we present the SDCs $f_{i}|_{i=0,1,2,3,4}$ defined in Eq.~(\ref{dsigmadcos}) for eleven different values of $\vert\cos\theta\vert$. It has been found that the numerical results of the SDCs $f_{i}|_{i=0,1,2,3,4}$ remain consistent whether we employ 10-digit or 20-digit precision for each Feynman integral family. In Table~\ref{SDCsfi1}, there are five non-logarithmic terms in $f_2$. The first three terms correspond to the vacuum polarisation and renormalisation contributions from the massless quark, charm quark and bottom quark, respectively. The fourth term corresponds to the light-by-light contributions from the massless quark (labeled as $lbl$). The fifth term represents all the other contributions, including the light-by-light contributions from charm quark and bottom quark. To check our calculation, we compare our numerical results at $\vert\cos\theta\vert=0.999$ with the values presented in Table B.1 in the supplementary material of Ref.~\cite{Sang:2023liy}. From Table~\ref{SDCsfi1}, we have $\pi f_{1}/f_{0}=-10.78$, which is exactly the same as that of Ref.~\cite{Sang:2023liy}. Taking the inputs $n_f=4$, $n_L=3$, $n_H=1$, $n_M=0$, and $lbl=0$, we have $\pi^2 f_{2}/f_{0}=-130.51-22.46L_{\mu}-51.18L_{\mu_{\Lambda}}$, which is consistent with that of Ref.~\cite{Sang:2023liy}. The light-by-light contributions are about 0.05\% of the NNLO predictions, which are indeed small and negligible as claimed in Ref.~\cite{Sang:2023liy}. Furthermore, in Table~\ref{SDCsfi1}, we observe that the numerical values of coefficients $f_{1}$ and $f_{2}$ are negative, resulting in negative corrections to LO predictions. In combination with $\alpha_s(\sqrt{s}/2)=0.1756$, the theoretical prediction for the process $e^+e^- \to J/\psi +J/\psi$ at NNLO is unphysical due to poor convergence behavior. However, by incorporating two additional parts of higher-order corrections, as illustrated in Table~\ref{SDCsfi2} and Table~\ref{SDCsfi3}, we are able to obtain a reasonable theoretical prediction. This is achieved by introducing positive corrections originating from $f_{3}$ and $f_{4}$.

\begin{table}[h]
\centering
\begin{tabular}{c c}
\hline
$\vert\cos\theta\vert$ & $f_{3}$ \\
\hline
$0.193$ & $92.0656+12.3660L_{\mu}+28.9002L_{\mu_{\Lambda}}$ \\
\hline
$0.402$ & $117.4510+15.9175L_{\mu}+36.9363L_{\mu_{\Lambda}}$ \\
\hline
$0.601$ & $178.1131+24.2128L_{\mu}+56.0634L_{\mu_{\Lambda}}$ \\
\hline
$0.698$ & $238.4933+32.2766L_{\mu}+75.0372L_{\mu_{\Lambda}}$ \\
\hline
$0.800$ & $359.3808+48.0764L_{\mu}+112.9460L_{\mu_{\Lambda}}$ \\
\hline
$0.849$ & $464.6776+61.6160L_{\mu}+145.9294L_{\mu_{\Lambda}}$ \\
\hline
$0.902$ & $669.2768+87.6152L_{\mu}+209.9949L_{\mu_{\Lambda}}$ \\
\hline
$0.922$ & $790.2901+102.8788L_{\mu}+247.8839L_{\mu_{\Lambda}}$ \\
\hline
$0.951$ & $1068.4986+137.7803L_{\mu}+334.9846L_{\mu_{\Lambda}}$ \\
\hline
$0.975$ & $1458.9501+186.4323L_{\mu}+457.1819L_{\mu_{\Lambda}}$ \\
\hline
$0.999$ & $1557.3600+196.9479L_{\mu}+487.0818L_{\mu_{\Lambda}}$ \\
\hline
\end{tabular}
\caption{The SDCs $f_{3}$ for eleven different values of $\vert\cos\theta\vert$. Here, $n_f=5$, $n_L=3$, $n_H=1$, $n_M=1$, $lbl=1$, $L_{\mu}=\ln\frac{\mu_R^2}{m_c^2}$, and $L_{\mu_{\Lambda}}=\ln\frac{\mu_{\Lambda}^2}{m_c^2}$.}  \label{SDCsfi2}
\end{table}

\begin{table}[h]\small
\centering
\begin{tabular}{c c}
\hline
$\vert\cos\theta\vert$ & $f_{4}$ \\
\hline
$0.193$ & \hspace{-0.6em} $209.5396+56.1687L_{\mu}+131.4701L_{\mu_{\Lambda}}+17.6318L_{\mu}L_{\mu_{\Lambda}}
+3.7722L_{\mu}^2+20.6262L_{\mu_{\Lambda}}^2$ \\
\hline
$0.402$ & \hspace{-0.6em} $265.3937+71.6562L_{\mu}+166.7230L_{\mu_{\Lambda}}+22.5346L_{\mu}L_{\mu_{\Lambda}}
+4.8556L_{\mu}^2+26.1955L_{\mu_{\Lambda}}^2$ \\
\hline
$0.601$ & \hspace{-0.6em} $401.8986+108.6657L_{\mu}+252.5597L_{\mu_{\Lambda}}+34.2039L_{\mu}L_{\mu_{\Lambda}}
+7.3860L_{\mu}^2+39.7032L_{\mu_{\Lambda}}^2$ \\
\hline
$0.698$ & \hspace{-0.6em} $540.8027+145.5020L_{\mu}+339.6221L_{\mu_{\Lambda}}+45.7797L_{\mu}L_{\mu_{\Lambda}}
+9.8459L_{\mu}^2+53.3639L_{\mu_{\Lambda}}^2$ \\
\hline
$0.800$ & \hspace{-0.6em} $824.0584+219.2561L_{\mu}+517.0743L_{\mu_{\Lambda}}+68.9077L_{\mu}L_{\mu_{\Lambda}}
+14.6655L_{\mu}^2+81.1595L_{\mu_{\Lambda}}^2$ \\
\hline
$0.849$ & \hspace{-0.6em} $1074.4249+283.4970L_{\mu}+673.7844L_{\mu_{\Lambda}}+89.0306L_{\mu}L_{\mu_{\Lambda}}
+18.7958L_{\mu}^2+105.6884L_{\mu_{\Lambda}}^2$ \\
\hline
$0.902$ & \hspace{-0.6em} $1566.1342+408.3217L_{\mu}+981.5335L_{\mu_{\Lambda}}+128.1166L_{\mu}L_{\mu_{\Lambda}}
+26.7267L_{\mu}^2+153.8510L_{\mu_{\Lambda}}^2$ \\
\hline
$0.922$ & \hspace{-0.6em} $1858.9494+482.1512L_{\mu}+1164.8197L_{\mu_{\Lambda}}+151.2324L_{\mu}L_{\mu_{\Lambda}}
+31.3829L_{\mu}^2+182.5361L_{\mu_{\Lambda}}^2$ \\
\hline
$0.951$ & \hspace{-0.6em} $2536.2604+651.8733L_{\mu}+1588.3677L_{\mu_{\Lambda}}+204.3721L_{\mu}L_{\mu_{\Lambda}}
+42.0295L_{\mu}^2+248.8244L_{\mu_{\Lambda}}^2$ \\
\hline
$0.975$ & \hspace{-0.6em} $3491.4667+890.0966L_{\mu}+2186.5424L_{\mu_{\Lambda}}+278.9239L_{\mu}L_{\mu_{\Lambda}}
+56.8706L_{\mu}^2+342.4140L_{\mu_{\Lambda}}^2$ \\
\hline
$0.999$ & \hspace{-0.6em} $3766.2171+950.1359L_{\mu}+2354.0053L_{\mu_{\Lambda}}+297.1657L_{\mu}L_{\mu_{\Lambda}}
+60.0783L_{\mu}^2+367.9209L_{\mu_{\Lambda}}^2$ \\
\hline
\end{tabular}
\caption{The SDCs $f_{4}$ for eleven different values of $\vert\cos\theta\vert$. Here, $n_f=5$, $n_L=3$, $n_H=1$, $n_M=1$, $lbl=1$, $L_{\mu}=\ln\frac{\mu_R^2}{m_c^2}$, and $L_{\mu_{\Lambda}}=\ln\frac{\mu_{\Lambda}^2}{m_c^2}$.}  \label{SDCsfi3}
\end{table}

\begin{figure}[htbp]
\centering
\includegraphics[width=0.8\textwidth]{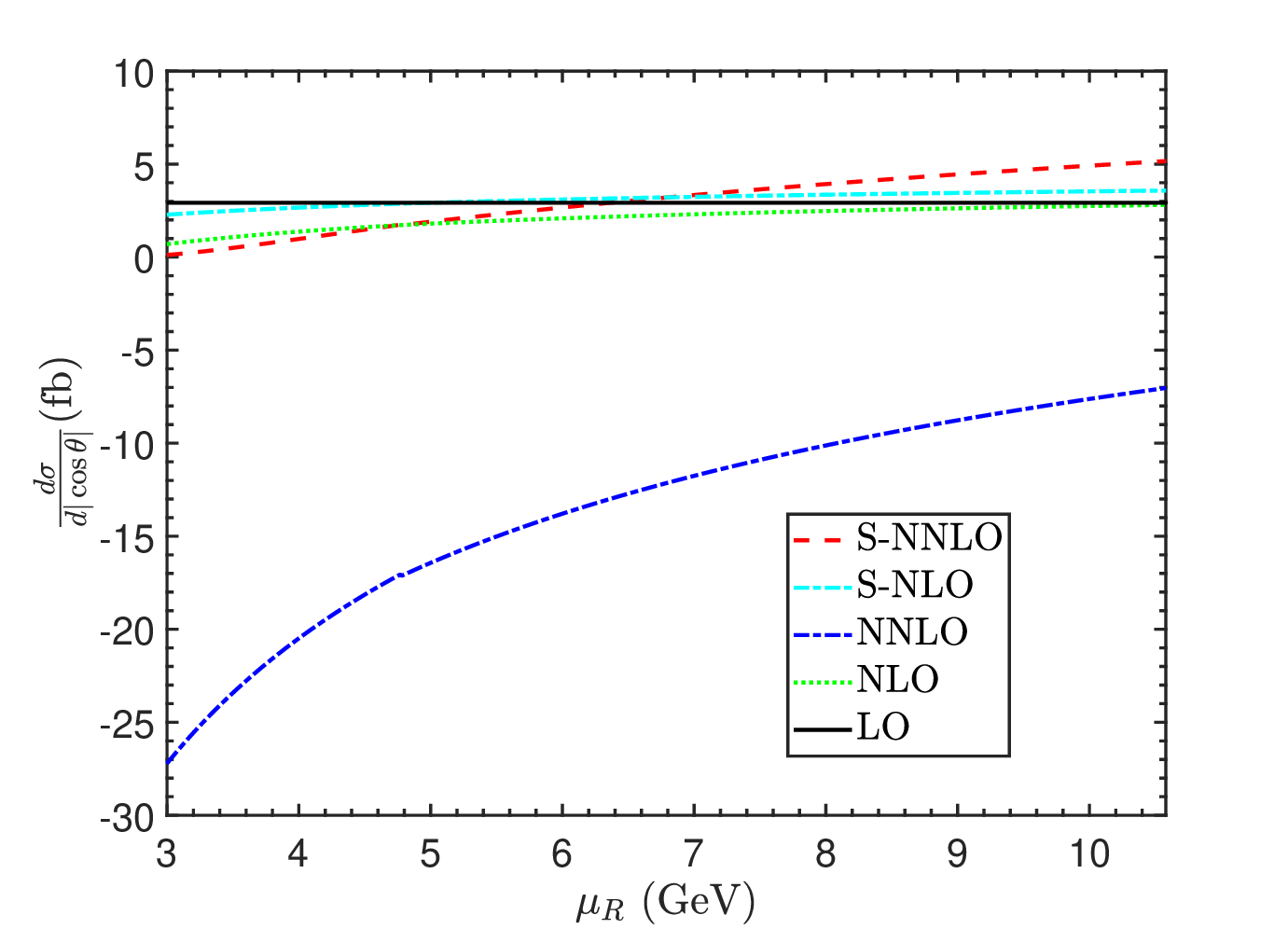}
\caption{The $\mu_R$ dependence of the differential cross section of $e^+e^- \to J/\psi + J/\psi$ at various perturbative order, and $\vert\cos\theta\vert=0.800$.} \label{urdependence8}
\end{figure}

In Fig.~\ref{urdependence8}, we illustrate the dependence of the renormalization scale $\mu_R$ on the differential cross section for the process $e^+e^- \to J/\psi + J/\psi$ at various perturbative orders with $\vert\cos\theta\vert=0.800$. It can be seen from the figure that the NNLO prediction exhibits the largest $\mu_R$ dependence among all the results. It can also be found that the prediction based on the square of NNLO amplitude (denoted as S-NNLO\footnote{Such a result is divergence free and gauge invariant. \label{denotedSNNLO}}) displays a more pronounced sensitivity to variations in $\mu_R$ when compared to both the NLO prediction and the prediction based on the square of NLO amplitude (denoted as S-NLO$^{\ref{denotedSNNLO}}$). The predictions based on S-NLO and S-NNLO closely approximate the LO prediction in the region $\mu_R\sim6.5$ GeV. This observation points to a notable convergence behavior in this particular region. 

\begin{figure}[htbp]
\centering
\includegraphics[width=0.8\textwidth]{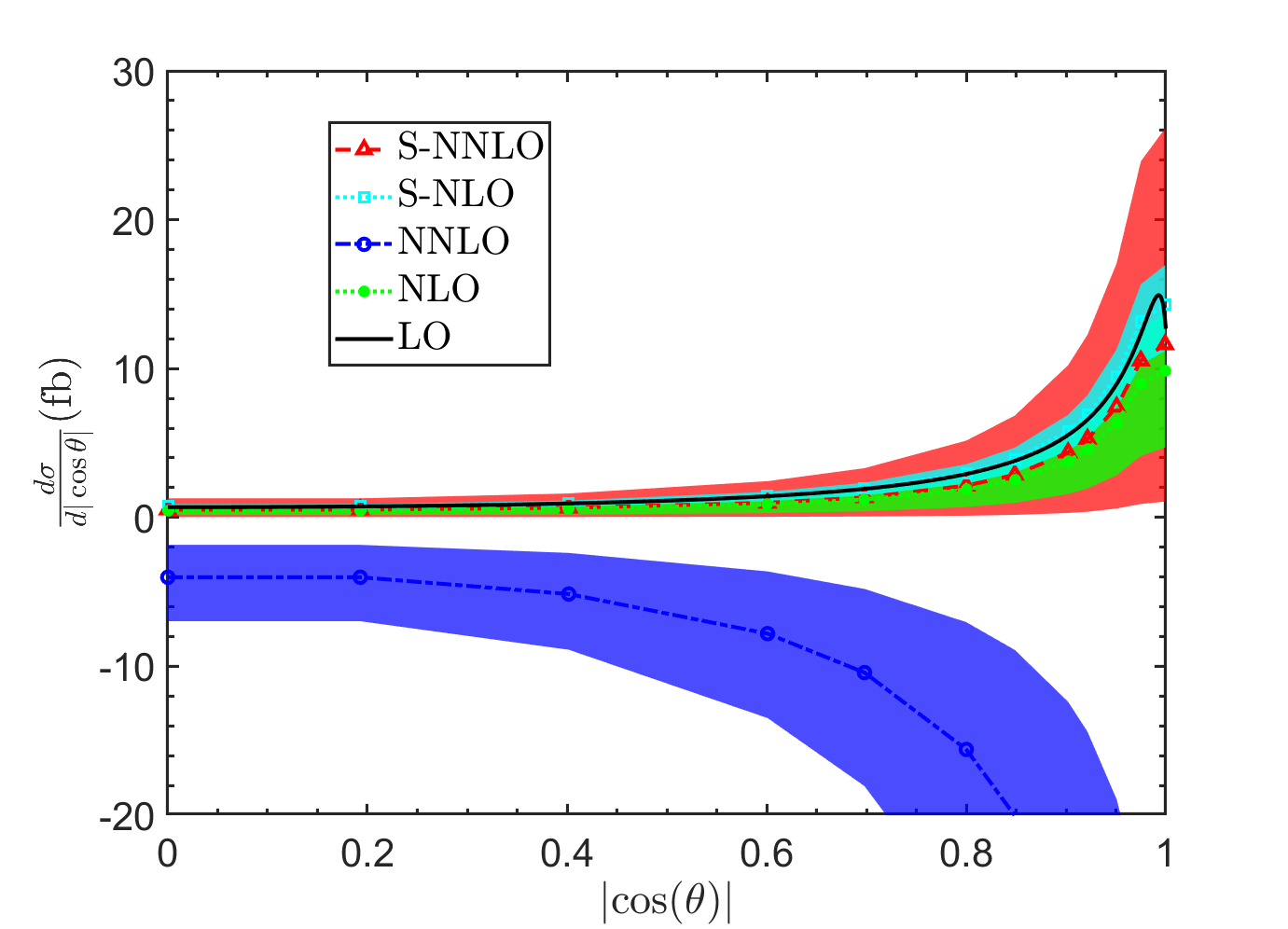}
\caption{The differential cross section of $e^+e^- \to J/\psi + J/\psi$ as function of $\vert\cos\theta\vert$ at various perturbative order, and the bands are obtained by varying the renormalization scale $\mu_R$ within the range of $[2m_c, \sqrt{s}]$.} \label{dsgdcos}
\end{figure}

In Fig.~\ref{dsgdcos}, we present the differential cross section for $e^+e^- \to J/\psi + J/\psi$ as a function of $\vert\cos\theta\vert$ at various perturbative orders. The central values are calculated with $\mu_R=\sqrt{s}/2$, and the bands represent the associated uncertainties arising from variation in $\mu_R$ within the range $[2m_c, \sqrt{s}]$. Fig.~\ref{dsgdcos} highlights the following observations: 1) The NNLO prediction yields a negative value with substantial uncertainty. 2) The prediction based on S-NNLO exhibits larger uncertainty compared to both the NLO prediction and the prediction based on S-NLO. The prediction based on S-NNLO covers the LO prediction, NLO prediction, and the prediction based on S-NLO in whole range. 3) The prediction based on S-NNLO is more reasonable than the NNLO prediction within the NRQCD factorization framework.

\section{Summary} \label{III}

In summary, we have computed the NNLO QCD corrections for the production of $J/\psi+J/\psi$ in $e^+e^-$ annihilation at a center-of-mass energy of $\sqrt{s}=10.58$ GeV. The numerical results of integrated cross section\footnote{The integrated cross section can be approximated using the trapezoidal rule~\cite{Ueberhuber:1997bk} with the results in Tables~\ref{SDCsfi1},~\ref{SDCsfi2}, and~\ref{SDCsfi3}. In other words, the trapezoidal rule is expressed as follows: $\int_{x_1}^{x_2}f(x)dx=\frac{f(x_2)+f(x_1)}{2}(x_2-x_1)$. We take the quantity $\frac{|f(x_2) - f(x_1)|}{2} \cdot (x_2 - x_1)$ as the absolute value of its uncertainty.} of $e^+e^- \to J/\psi + J/\psi$ with three typical renormalization scales $\mu_R$ at different perturbative orders are shown in Table~\ref{mur1}. It can be seen that the prediction at NNLO becomes negative due to the poor convergence of perturbative expansion. However, based on the square of NNLO amplitude, the theoretical prediction of the cross section becomes reasonable. The similar situation could be found in the case of NNLO result for $J/\psi\rightarrow e^+ e^-$ where only the square of NNLO amplitude treatment provides reasonable estimation.  

\begin{table}[h]
\centering
\begin{tabular}{c c c c c c }
\hline
$\sigma (fb)$ & LO & NLO & NNLO & S-NLO & S-NNLO \\
\hline
 $\mu_R=2m_c$ & $2.29$ & $0.61$ & $-21.10$ & $1.83$ & $0.12$  \\
\hline
 $\mu_R=\sqrt{s}/2$ & $2.29$ & $1.54$ & $-11.97$ & $2.37$ & $1.76$  \\
\hline
 $\mu_R=\sqrt{s}$ & $2.29$ & $2.25$ & $-5.27$ & $2.84$ & $4.17$  \\
\hline
\end{tabular}
\caption{ The integrated cross section of $e^+e^- \to J/\psi + J/\psi$ with three typical renormalization scales $\mu_R$ at various perturbative accuracy.}  \label{mur1}
\end{table}

Finally, we have derived the theoretical prediction based on S-NNLO for the process $e^+e^- \to J/\psi+J/\psi$ at the $B$ factories, i.e.,
\begin{eqnarray}
\sigma_{\rm S-NNLO}&=& 1.76^{+2.41+0.25}_{-1.64-0.25} \nonumber \\
&=&1.76^{+2.42}_{-1.66} ~({\rm fb}), \label{result}
\end{eqnarray}
where the center value is obtained by taking $m_c=1.5$ GeV\footnote{It is worth noting that our calculation does not cover different choices of charm quark pole mass, it is due to the time-consuming nature of the IBP reduction within the auxiliary mass approach employed in AMFlow. Thus, the actual uncertainty may be somewhat larger than that shown in Eq.~(\ref{result}).}, $\mu_R=\sqrt{s}/2$, and $\mu_\Lambda=1$ GeV. The first uncertainty arises from the variation of $\mu_R$ within the range $[2m_c, \sqrt{s}]$, and the second uncertainty is attributed to the method used to estimate the integrated cross section from the differential cross section. It should be pointed out that the center value of our prediction (1.76 fb) is lower than the value (2.13 fb) presented in Table.~I of Ref.~\cite{Sang:2023liy}, and our result exhibits a more pronounced dependence on the choice of $\mu_R$. Furthermore, we find that the light-by-light contributions are indeed small and negligible as claimed in Ref.~\cite{Sang:2023liy}. Both of these theoretical predictions are lower than the upper limit of $\sigma[e^+e^- \to J/\psi +J/\psi] \times \mathcal{B}_{>2} < 9.1~\mathrm{fb}$.  Our result for total cross section and differential cross section could be compared with precise experimental measurement in future at the $B$ factories.

\hspace{2cm}

\noindent {\bf Acknowledgments:} We would like to thank Yan-Qing Ma for helpful suggestions and Jichen Pan for useful discussions. This work was supported by the National Natural Science Foundation of China with Grant Nos. 12247129, 11975242, and 12135013. It was also supported in part by National Key Research and Development Program of China under Contract No. 2020YFA0406400.

\hspace{2cm}

\bibliographystyle{JHEP}
\bibliography{nrqcd}

\end{document}